\documentclass[11pt]{article}
\usepackage{times}
\usepackage{geometry}
\usepackage[utf8]{inputenc}
\usepackage{enumitem,amssymb}
\usepackage{graphicx}
\usepackage{fancyhdr}
\usepackage{aas_macros}
\usepackage{mdframed} 
\usepackage{amsmath}

\usepackage[authoryear]{natbib}
\setcitestyle{authoryear,open={(},close={)}}

\newcommand{\arcmin}{\ensuremath{^\prime}}
\newcommand{\Hi}{H~{\sc i}}
\newcommand{\Hii}{H~{\sc ii}}

\newcommand{\MHz}{\ensuremath{\textrm{ MHz}}}

\providecommand{\sorthelp}[1]{}

\mdfdefinestyle{theoremstyle}{
innertopmargin=\topskip,}
\mdtheorem[style=theoremstyle]{lrptextbox}{}

\title{Canadian Investigations of the Interstellar Medium}
\author{Alex S. Hill (UBC) \and
Jan Cami (Western University) \and
Laura Fissel (Queen's University) \and
Tyler Foster (Brandon U) \and
Gilles Joncas (U Laval) \and
Lewis Knee (NRC/HAA) \and
Roland Kothes (NRC/HAA) \and
Tom Landecker (NRC/HAA) \and
Tim Robishaw (NRC/HAA) \and
Erik Rosolowsky (U Alberta) \and
Samar Safi-Harb (U Manitoba) \and
Jennifer West (U Toronto) \and
Trey V. Wenger (NRC/HAA)}

\begin{document}

\setcounter{page}{0}

\maketitle

\section*{Executive Summary}

The interstellar medium (ISM) mediates galactic evolution as the reservoir of material for future star formation and the repository of energy and matter output by stellar processes. It consists of matter (both gas and dust), magnetic fields, and relativistic particles in rough equipartition in energy density. Understanding this equilibrium requires observing the discrete thermally-stable phases as well as tracing the phase transitions and turbulence that mix the phases and produce unstable gas. The heating, cooling, and ionization of the ISM trace the flow of radiation through galaxies, so the impact of the ISM must be understood to interpret observations of most astrophysical phenomena.

Canadians have played leading roles in ISM science for decades. The Canadian user community for the Dominion Radio Astrophysical Observatory (DRAO) Synthesis Telescope designed and completed the Canadian Galactic Plane Survey (CGPS). The CGPS identified a wealth of small-scale structure in H I emission as well as self-absorption and in the structure of polarized emission. These observations demonstrated that no phase of the ISM, including the transition from atomic gas to star formation, can be understood in isolation. Canadians have also played leading roles in the characterization of dust with Planck and balloon-borne telescopes. Canadians have also used pulsar scintillometry to measure structure in the ISM at the smallest scales, below 1 AU.

The 2020s offer many opportunities for ISM science in Canada. A major but cost-effective upgrade to the Synthesis Telescope with broadband (400--1800 MHz) single-pixel feeds would enable broadband polarimetry as well as wide-area, arcminute surveys of radio recombination line emission tracing ionized gas that cannot be observed in any other way. The next generation of balloon-borne telescopes will investigate magnetic fields and dust properties in a wide range of regions. Large single dishes remain essential for our understanding of the diffuse structure which characterizes the ISM. The Green Bank Telescope is the only facility available to Canadians capable of observing the range of molecular transitions from $1-116$~GHz. Very long baseline interferometry (VLBI) capability enables parallax measurements of pulsars and masers, needed for measuring the structure of the Galaxy and the ISM within it, and for further progress in scintillometry. Canadian ISM astronomers will continue to participate in cosmological experiments including CHIME and CHORD. Protecting quiet radio frequency interference environments at radio observatories will be ever more important as broadband observations are ever more central to ISM spectral line and continuum science. Computational capability is essential both for numerical work and for handling the observational data.

\clearpage

\section{Introduction}

The evolution of galaxies is driven primarily by mergers and accretion, feedback from galactic nuclei, and star formation feedback. The interstellar medium (ISM) mediates all of these processes. It consists of matter (both gas and dust), magnetic fields, and relativistic particles (cosmic rays) in rough equipartition in energy density \citep[see review by][]{Ferriere:2001}. The basic modern picture of the gas phases dates to \citet{McKee:Ostriker:1977}, who formulated a model in which supernova explosions and radiative heating balanced by atomic cooling regulate a three-``phase'' ISM in approximate thermal pressure equilibrium. Stars form out of the dense gas, but the dense gas is presumably fed by lower-density gas through unknown mechanisms. The enormous range of scales and physical processes in the ISM requires a diverse set of observations. This also means that the ISM, as a diffuse component of the Galaxy that affects every sightline, must be understood as a foreground for most other astronomical observations, but it can be observed in concert with facilities designed primarily for other scientific objectives.


Canada has played an outsized role in ISM science for decades. The Dominion Radio Astrophysical Observatory (DRAO)'s Synthesis Telescope (ST) was the first instrument to provide an arcminute view of \Hi\ emission, thus revealing the complexity of the distribution of neutral gas \citep{Landecker:Higgs:1980}. In the 1990s, the ST user community came together to plan and complete the Canadian Galactic Plane Survey (CGPS), a survey which included \Hi, diffuse polarized and total continuum at 1420~MHz, and total continuum at 408~MHz. The CGPS demonstrated that no phase of the ISM can be understood in isolation. The CGPS found the first evidence of widespread cold \Hi\ seen in self-absorption, the first stage of star formation \citep{Gibson:Taylor:2005}, pioneered the Faraday rotation measure (RM) grid technique \citep{Brown:Taylor:2001,Brown:Taylor:2003}, and used diffuse polarized emission to identify pc-scale turbulent structures in the ISM \citep{Landecker:Reich:2010}. The DRAO \Hi\ Intermediate Galactic Latitude Survey (DHIGLS; \citealt{Blagrave:Martin:2017}) has produced measurements of the three-dimensional structure of cold gas at intermediate latitudes. Canadians were leading participants in {\em Planck}'s dust and insterstellar magnetic field efforts \citep{planck2014-XIX}. In addition, the Balloon-borne Large Aperture Sub-mm Telescope for Polarimetry (BLASTPol), which had significant leadership from the University of Toronto and UBC, made large-scale dust polarimetry observations toward molecular and translucent clouds in three sub-mm bands.  Combining BLASTPol and {\em Planck} polarization observations shows a flat sub-mm polarization spectrum \citep{Gandilo:Ade:2016,Ashton:Ade:2018}, which is in tension with predictions of many standard two-component dust models (e.g.~\citealt{Draine:Fraisse:2009}). Moreover, the Canadian WIDAR correlator on the Karl G.\ Jansky Very Large Array (VLA) has enabled a wide variety of ISM science for global users.

In this white paper, we concentrate on the areas of opportunity in the 2020s for Canada in the non-star forming ISM. Canadian expertise and facilities lead us to three core science themes: i) the ionized ISM, ii) interstellar dust, and iii) diffuse molecular gas and the onset of star formation.
White paper E024 by Rosolowsky et al.\ focuses on galactic evolution in combination with star formation, while white paper E025 by di~Francesco et al.\ focuses on star formation itself. Magnetic fields are a crucial component of the ISM and cannot be understood without understanding the gas because all observational tracers of magnetic fields are intertwined with the gas; white paper E064 by West et al.\ covers cosmic magnetism.

\section{Science areas} \label{sec:science}

\subsection{Galactic structure}

The ISM is organized by, and is a key tracer of, the structure of the Galaxy. The morphological and chemical structure of the Galaxy is a vital constraint on theories of Milky Way formation and chemodynamical evolution. Dust extinction in the Galactic plane limits most tracers of Galactic structure to radio wavelengths. Common radio tracers are \Hi, CO, and high mass star forming regions. Although \Hi\ and CO permeate the Galaxy and show evidence of spiral structure in the Milky Way \citep[e.g.,][]{Koo:2017,Dame:2011}, it is difficult to estimate distance to the source of such emission due to the complicated dynamics of the Galaxy, especially within the solar orbit. \Hii\ regions are a classic tracer of structure in disk galaxies due to their association with star-forming spiral arms.
Distances to such objects are from maser parallax, by estimating the spectral type of the ionizing star(s), or by assuming a model of Galactic rotation. Kinematic distances are the most easily obtained, but can be wildly inaccurate due to non-circular motions as well as the near/far distance ambiguity in the inner Galaxy \citep[see][]{Wenger:2018}. Surveys in search of new Galactic \Hii\ regions are filling-in the unexplored regions of the Galaxy, especially in the Southern sky \citep[e.g.,][]{Wenger:2019}, and new techniques are being developed to create a more accurate picture of the Milky Way's velocity field \citep[e.g.,][]{Tchernyshyov:2017}.

\subsection{Multi-phase structure of the gas}

The ISM is inherently multi-scale with a wide range of important physical processes, with power on spatial scales spanning as much as twelve orders of magnitude in just the ionized gas \citep{Chepurnov:Lazarian:2010,Lee:Lee:2019}. 
One of the key theoretical questions in star formation is whether the large scales or small scales are the controlling steps in sustaining star formation. A step change in our understanding of galactic star formation will come from simultaneously including gravitational collapse and star formation feedback within giant molecular clouds on sub-pc scales while including galactic effects like rotation and hydrostatic support on $\sim 50 \textrm{ kpc}$ scales. Diffuse ISM work has shown that adaptive mesh Eulerian simulations with $\sim 5 \textrm{ pc}$ peak resolution are required to achieve a converged three-phase ISM \citep{deAvillez:Breitschwerdt:2004,Hill:Joung:2012,Kim:Ostriker:2018}; current simulations of full galactic disks remain an order of magnitude away from this \citep{Koertgen:Banerjee:2019} or use smoothed particle hydrodynamics \citep{Dobbs:Pettit:2018,Dobbs:Rosolowsky:2019}, which is not optimized for capturing shocks that produce the hot phase of the ISM. Resolving gravitational collapse somewhat realistically requires another order of magnitude in spatial resolution \citep{Ibanez-Mejia:MacLow:2017}. Therefore, numerical work in the 2020s will require the computational resources to bridge from the sub-pc scales on which star formation proceeds to the galactic scales on which the ISM and the star formation environment are organized.
Whereas theoretical and numerical work can incorporate a range of physical processes, observations tend to be sensitive to a limited set of phases. Therefore, we must attempt to synthesize a complete picture of the multi-phase ISM from diverse observations. In the rest of this section, we discuss observational approaches to different components of the ISM in turn.

\subsubsection{Ionized gas}

A significant fraction of interstellar hydrogen is ionized. Ionizing radiation from massive OB~stars creates locally-ionized \Hii\ regions. These stars and nebulae are short-lived ($\lesssim 10$ Myr), so \Hii\ regions reveal the locations of \textit{current} massive star formation. In other galaxies, \Hii\ regions are primarily found in star-forming spiral arms, so \Hii\ regions are the classic tracer of spiral structure in the Milky Way. Dust extinction prevents the optical identification (i.e., via H$\alpha$ emission) of Galactic \Hii\ regions. On-going surveys have discovered thousands of new nebulae by their radio recombination line (RRL) emission \citep[e.g.,][]{Wenger:2019}, and soon we will have a catalog of the positions, radial velocities, and RRL properties of all Galactic O~star \Hii\ regions.
Although \Hii\ regions produce bright H$\alpha$ and RRL emission due to their compactness, $90\%$ of the H$^+$ gas mass and $40-80\%$ of the H$\alpha$ luminosity in the Milky Way and other star-forming galaxies is in the extensive, diffuse, low surface brightness component called the warm ionized medium (WIM; \citealt{Haffner:Dettmar:2009}). The power requirement indicates that the WIM must be closely tied to star formation. Mapping the Milky Way's WIM is one of the last great challenges in developing a comprehensive view of its interstellar gas, and penetrating into the different environments in the inner Galaxy will allow identification of the widespread ionization sources. The WIM was first discovered through free-free absorption \citep{Hoyle:Ellis:1963}, but many observational tracers are now used.

{\bf Pulsar dispersion} provides a direct probe of the free electron column density. In combination with the distance to the pulsar, this provides the space-averaged mean electron density; with many pulsars with known distances, one can construct a model of the free electron distribution and spiral structure of the Galaxy \citep{Taylor:Cordes:1993,Cordes:Lazio:2002,Yao:Manchester:2017}. These models are in turn used to estimate the distance to pulsars for pulsar science and for characterizing fast radio bursts \citep{2019Natur.566..230C}, but using these distances for studies of the ISM is circular. Therefore, DM-independent distances are required. The gold standard method is parallax, either in the radio using very long baseline interferometry (VLBI) techniques \citep[eg][]{Deller:Goss:2019} or in the optical, most recently with {\em Gaia} \citep{Jennings:Kaplan:2018}, but the sampling remains sparse. Many more pulsar parallax measurements and improved estimates with other techniques like \Hi\ absorption combined with a Galactic rotation model will significantly improve our understanding of the distribution of free electrons.

{\bf Optical emission lines} provide the bulk of our understanding of the large-scale distribution and physical conditions in the WIM. Two imaging surveys, the Southern H-Alpha Sky Survey Atlas (SHASSA) and the northern Virginia Tech Sky Survey (VTSS), have produced spectacular arcminute-scale views of the WIM across most of the sky but lack the sensitivity and spectral resolution to image the three-dimensional distribution of the gas. The Wisconsin H-Alpha Mapper (WHAM; \citealt{Haffner:Reynolds:2003,Haffner:Reynolds:2010}) has provided kinematically-resolved H$\alpha$ maps of the entire sky at $1^\circ$ resolution as well as wide-area maps of other optical emission lines to constrain gas temperature, ionization state, and the ionizing radiation field \citep{Madsen:Reynolds:2006,Hill:Benjamin:2014}. WHAM is sensitive to an emission measure $\textrm{EM} \equiv \int n_e^2 \, ds \approx 0.2 \textrm{ pc cm}^{-6}$; this sensitivity is largely limited by atmospheric emission lines and is unmatched by any other emission WIM probe now or in the foreseeable future. The three main limitations of WHAM are the poor angular resolution, extinction effects, and a spectral window which requires multiple observations to observe the Magellanic System and other extraplanar gas. A higher-resolution successor to WHAM with wider spectral coverage should be an American priority for the coming decade, but Canada is well-positioned to complement WHAM with fundamentally new radio capabilities to observe other tracers of the WIM.

{\bf Radio recombination lines} (RRLs) are extraordinarily faint (the best RRL surveys, employing stacking of many lines with modern broadband telescopes, achieve an EM sensitivity of $\sim 150 \textrm{ pc cm}^{-6}$), but in the Galactic plane, the intrinsic emission measures from the WIM are often this high \citep{Madsen:Reynolds:2005,Krishnarao:Benjamin:2019} while the optical emission is obscured by dust. Moreover, the arcminute angular resolution that is possible with RRLs is very important in the disk. The state of the art in wide-field arcminute-scale RRL WIM surveys today is the ongoing Survey of Ionized Gas of the Galaxy \citep[SIGGMA;][]{Liu:Anderson:2019}, part of the Galactic Arecibo L-band Feed Array (GALFA) family of surveys; and the Green Bank Telescope Diffuse Ionized Gas Survey \citep[GDIGS;][]{Luisi:2018}. SIGGMA will observe two narrow strips across the Galactic plane accessible to the telescope and has released a 6-arcminute RRL survey of the Inner Galaxy from $32^\circ \leq \ell\leq 70^\circ$, $|b|\leq 1.5^\circ$. Being single-dish, SIGGMA can image extended spatial structures down to the resolution limit of the telescope. However, as an Arecibo survey, SIGGMA is limited to declinations $0^\circ \lesssim \delta \lesssim +40^\circ$, missing much of the Galactic plane. GDIGS is mapping the inner Galactic plane unobservable by Arecibo ($-5^\circ \leq \ell \leq 32^\circ$, $|b| \leq 0.5^\circ$). The upgraded ST (see white paper E080) will achieve comparable sensitivity to SIGGMA but with spatial resolution $\approx 2$ times better and with the ability to image any field in the northern sky. These projects, combined with the \Hii\ region catalogs from recent \Hii\ region discovery surveys, will allow us to investigate how ionizing photons leak out of \Hii\ regions and power the WIM.

{\bf Faraday rotation} probes the convolution of the magnetic field and the ionized gas by providing the Faraday depth $\phi \propto \int n_e \vec{B} \cdot d\vec{s}$. The advent of wide-bandwidth radio feeds and spectrometers in combination with computational capacity to apply the Faraday synthesis technique \citep{Burn:1966,Brentjens:deBruyn:2005} has revolutionized investigations of magnetized gas by making it possible to separate emission components at multiple ``Faraday depths'', analogous to the ability of spectral line observations to identify gas at multiple velocities. The ISM causes Faraday rotation both in background point sources (primarily active galactic nuclei and pulsars) and in diffuse synchrotron emission, which mostly originates in our Galaxy. With many background polarized sources, this can be used to map the structure of the magnetoionized foreground gas in the Galactic disk \citep[eg][]{Brown:Taylor:2003} and halo \citep{Betti:Hill:2019}. We are just beginning to understand how to use Faraday depth information from diffuse emission to construct a three-dimensional picture of the magnetoionized ISM \citep{Dickey:Landecker:2019}. The exciting prospects for these techniques in the 2020s are discussed in WP E080 by West et al.

\subsubsection{Neutral gas}

The bulk of the mass the ISM is in neutral atomic hydrogen both in warm (WNM) and cold (CNM) states. The \Hi\ 21~cm emission line provides the bulk of our information on the physical conditions and distribution of H$^0$ in the Galaxy. Between the CGPS and its counterparts, the Southern Galactic Plane Survey (SGPS) and the VLA Galactic Plane Survey (VGPS), the entire Galactic disk has been mapped with arcminute resolution. Meanwhile, single dish surveys have mapped the entire sky at $16'$ resolution to form the HI4PI survey \citep{BenBekhti:Floer:2016}.

The next step change in our understanding of neutral gas is likely to come from the incredibly wide fields made possible by phased array feeds (PAFs). ASKAP has observed the entire Small Magellanic Cloud in one exposure with $30''$ resolution \citep{McClure-Griffiths:2018}, exceeding the quality of observations which took hundreds of hours using the Australia Telescope Compact Array \citep{Stanimirovic:Staveley-Smith:1999} in just 36~hours with a telescope without a dramatically larger collecting area. The GASKAP project \citep{Dickey:McClure-Griffiths:2013}, which has Canadian involvement, will survey the southern sky Galactic plane as well as the entire Magellanic System, including both galaxies and the Magellanic Stream and Leading Arm.
The two octave frequency coverage of the single pixel feeds on the proposed renewal of the ST is not possible with PAFs, but exploring PAF technology on the ST engineering testbed could lead the ST to being an outstanding northern hemisphere widefield \Hi\ and OH imaging machine.

\subsubsection{Molecular gas}

The role of molecular gas is discussed extensively in E024 and E025 focusing on star formation.  However, in context of the ISM, molecular gas is a major and often dominant mass reservoir. Molecular line transitions are required for cooling the ISM to $T\lesssim 10~\mathrm{K}$ and the enrichment and chemistry of molecular gas in the gas phase or on the surface of dust grains plays a major role in establishing the chemistry of stellar systems. While ALMA and the JCMT play a key role here at high frequencies, the centimetre and millimetre wavebands are essential for a complete understanding of the ISM.  The Green Bank Telescope is the only single-dish facility that Canadians can access covering the range of 1 to 115 GHz where many key molecules have their brightest transitions. In particular, the multi-pixel feeds on the GBT have enabled recent studies of the chemical complexity of the ISM by Canadian-led teams conducting large programs \citep{friesen17, keown19}.

A particular subset of important molecular species are large aromatics such as polycyclic aromatic hydrocarbons \citep[PAHs; see][]{2008ARA&A..46..289T}, fullerenes \citep{Cami:APN_formation}, and the putative carriers of the diffuse interstellar bands \citep[DIBs; see e.g.][]{DIBconf2014} -- a set of about 600 unidentified interstellar absorption bands from the near-UV to the near-IR whose carrier molecules remain unknown -- with the exception of C$_{60}^+$ that has been identified with several DIBs \citep{Campbell:C60+DIBs}. These large aromatic species lock up $\sim$15--20\% of the cosmic carbon, and play key roles in heating and cooling, and thus large-scale processes such as star and planet formation. Canadian astronomers have a leading role in state of the art research on the molecular properties of these species and their role in large-scale processes (see also E063 by Cami et al.). 

\subsubsection{Zeeman splitting measurements and the atomic-molecular transition}

Radio measurements of the Zeeman effect provide an in-situ method for characterizing the Galactic magnetic field on multiple scales: diffuse 21-cm \Hi\ emission can probe the large-scale field; 18-cm OH transitions (thermal and masing) can trace the field inside molecular clouds and star-forming regions; and RRLs can probe the field in both the diffuse ionized medium and compressed photodissociation regions. Such measurements have been of limited profit in the past due to a combination of the long observing times required, the polarization conversion of incoming radio signals by the telescope (i.e., instrumental polarization; \citealt{Robishaw:Heiles:2018}), and the availability of backends with limited spectral bandwidth and resolution.  Many of these limitations will be lifted by the telescopes of the next decades: FAST, SKA, and ngVLA will increase available collecting area; FPGA and GPU technology are already being employed to produce full-Stokes spectral-line backends that allow for exquisite spectral resolution over gigahertz bandwidth; electromagnetic modeling software can now effectively allow for the characterization of instrumental polarization in radio telescopes \citep{Du:Landecker:2016}.  Large surveys of Zeeman-splitting in \Hi\ and OH emission are being planned on the upgraded DRAO John A.\ Galt 26-m, the Parkes 64-m, and the Effelsberg 100-m.  The long dwell times required for Zeeman detections on single-dish telescopes coupled with new wideband front- and backends will allow for simultaneous collection of all 18-cm OH transitions and dozens of RRLs guaranteeing the total-power detection of these transitions to levels seldom observed.  These deep, multi-line observations will facilitate the study of the atomic-molecular transition in molecular clouds and filaments.  An upgraded DRAO Synthesis Telescope (described below) will be designed for similar multi-line ISM studies at arc-minute resolution and will facilitate blind surveys for OH maser emission from star-forming regions and OH/IR stars and OH absorption towards \Hii\ regions and extragalactic sources.

\subsubsection{Dust polarization measurements}

Dust is ubiquitous across almost all phases of the ISM. Thermal radiation from dust grains aligned on average with their minor axes parallel to the local magnetic field can be used to trace magnetic field morphology across many phases of the ISM \citep{Andersson:Lazarian:2015}.  {\em Planck} all-sky polarization maps at 353\,GHz have led to significant advances in our understanding of magnetic fields in the diffuse ISM \citep{planck2014-XIX}.  However for polarization maps  {\em Planck} was generally sensitivity limited to a FWHM resolution of $\geq 10\arcmin$.

The BLASTPol balloon polarimeter produced extremely detailed maps of nearby molecular clouds \citep{Fissel:Ade:2016}.  The BLAST-TNG telescope will produce polarization maps of many more molecular clouds but also large maps of CNM emission, where it will be possible to compare dust polarization with Faraday rotation measures from the POSSUM survey with ASKAP map the 3D dimensional structure of the magnetic field.

\subsection{Inflow and high velocity clouds}

Galaxies like the Milky Way require a source of low-metallicity material to sustain their observed star formation histories. High velocity clouds (HVCs) -- clouds of gas with velocities inconsistent with Galactic rotation -- are a likely source of accreting material with sub-solar metallicity. However, it is unclear whether HVCs can cool, condense, and join the star forming reservoir in the disk. Arcminute scale wide field observations of HVCs will resolve the (magneto)hydrodynamic interactions as they pass into the outer Galactic disk. To date, such observations are uncommon, in part because the limited spectral window of the ST limits our ability to observe the disk and HVCs simultaneously with good spectral resolution. Forster et al (submitted to {\em Nature}) observed a grouping of high velocity clouds with the ST, and found that the \Hi-emitting gas interacts with and compresses the gas emitting polarized radio continuum. This is the first image of emission from magnetized HVCs and demonstrates the enormous potential of the ST to apply the multi-phase observations from the CGPS to extra-planar environments. In the southern sky, GASKAP's $0.5'$ observations will fundamentally change our understanding of the dynamical processes at work in the Magellanic System.

\subsection{Small-scale structure and scintillometry}

Small-scale inhomogenities in the ISM lead to interference patterns in radio signals from compact sources, primarily pulsars. These interference patterns produce characteristic changes in the spectra of pulsars on timescales of minutes, which can be inverted to probe interstellar inhomogenities on scales from AU down to thousands of km \citep{Hill:Stinebring:2005,Brisken:Macquart:2010}. It is unclear whether the observed features are due to discrete plasma structures \citep{Gwinn:2019} or corrugated plasma sheets \citep{Simard:Pen:2018}; distinguishing between these possibilities has implications for whether scintillometry is measuring the inner scale of turbulence and therefore the scale on which turbulence dissipates and heats the ISM. Retaining a VLBI capability is essential for further progress.

\subsection{Supernova remnants}

Supernova explosions are some of the most transformative events in our Universe and their remnants are fundamental to understanding the chemical enrichment, turbulence, and magnetism in galaxies, including our own Milky Way. Supernova remnants (SNRs) are a critical component of the ISM, injecting energy and compressing the magnetic field. SNR studies are a powerful tool to study the intrinsic properties of the explosion and the environment in which they are expanding. They have long been proposed to be the primary sources of high-energy cosmic rays
\citep[e.g.,][]{2001RPPh...64..429M}. Multi-wavelength studies, particularly high-energy observations, have shown that SNRs can accelerate electrons, and even protons in some cases, up to the knee of the cosmic ray spectrum \citep{1995Natur.378..255K,  2013Sci...339..807A} and compress magnetic fields up to milli-Gauss \citep[e.g.,][]{2007Natur.449..576U}. The question however remains whether protons, which make the bulk of cosmic rays, are efficiently accelerated in SNRs and Pulsar Wind Nebulae up to the knee of the cosmic ray spectrum \citep[e.g.,][]{2007ApJ...665L.131G} as well the mechanism for particle acceleration and cosmic ray escape in supernova shocks remains a puzzle. This is driving the next generation of X-ray and gamma-ray telescopes such as ATHENA \citep{2013arXiv1306.2335D} and the Cherenkov Telescope Array \citep{2019scta.book.....C}.  As well, state-of-the-art 3D simulations of SNRs are needed to compare the 3D structure and spectra of SNRs acting as efficient particle accelerators to observations across the electromagnetic spectrum \citep[e.g.,][]{2014ApJ...789...49F}. Canadians are involved in many aspects of this research from gamma-ray instrumentation \citep[][]{2015NIMPA.797...13H}, to key partnership in X-ray missions with international agencies through the Canadian Space Agency (ASTROSAT and Hitomi), to the development of Canada's flagship X-ray telescope, Colibr\`i (White Paper E048, Hoffman et al.), to multi-wavelength observations from radio to high-energy gamma-rays \citep{2016A&A...587A.148W, 2017ASSL..446....1K, 2019BAAS...51c.317S, 2017ApJ...836...23A, 2018A&A...612A...3H}, to modelling \citep{2017A&A...597A.121W, 2014ApJ...792..133F} and 3D numerical simulations performed on supercomputers \citep[e.g.,][]{2016sros.confE..92F}.

\subsection{As a cosmological foreground}

All observations of the Universe must see through the ISM of the Milky Way. Even at high Galactic latitudes, the ISM has a significant effect on observations across the electromagnetic spectrum. This is a challenge for extragalactic and cosmological observations and an opportunity for Galactic science. It has led to fruitful collaborations within the teams operating cosmological missions including the {\em WMAP}, {\em Planck}, and Canadian Hydrogen Intensity Mapping Experiment (CHIME) teams measuring free-free emission from the WIM, magnetized dust, and numerous other applications. In CHIME, the signal from redshifted \Hi\ emission at $400-800 \MHz$ (redshift $2.5 > z > 0.8$) is $\sim 10^{-5}$ times the intensity of the continuum emission (primarily synchrotron) from the Milky Way's ISM. Although the principle method for removing the CHIME foregrounds relies on the fact that any large feature in angle or frequency must be local and therefore requires relatively little understanding of the ISM, polarized emission Faraday rotated by the ISM or the ionosphere leaking into Stokes $I$ will have fine frequency structure and is likely to be $\sim 10$ times brighter than the cosmological signal. CHIME polarization data will be used to characterize the three-dimensional magnetoionized medium (see WP E081). CHIME will also be used to measure spectra of extended and compact structures in the Galaxy to understand the emission mechanisms.

\section{Resource requirements} \label{sec:resources}

Understanding of the wide ranges of physical conditions, scales, and temperatures in the diffuse ISM requires observations at a range of wavelengths and multi-scale computational work. {\bf The LRP should support the following resources as priorities for ISM science.} Most modern ISM science is limited by computational capabilities.

Both observational and numerical work on the ISM has considerable {\bf computational requirements}. As the link between star formation on small scales and Galactic-scale processes, numerical studies of the ISM require a multi-scale approach. Simulations which push the capabilities of Canadian high-performance computing capabilities are now routine globally. Compute Canada is an appropriate place to house general-purpose high performance computing such as is typically used for numerical work. The next generation of radio telescopes are fundamentally limited by computing capacity both in processing the data from the telescope and for long-term storage and analysis. SKA is considering a model in which advanced data processing is distributed across regional data centres; a similar approach may become necessary to handle data from other upcoming instruments. Science with phased array feeds in particular is currently limited more by computational capability than by hardware.

The {\bf renewed Synthesis Telescope's} (see white paper E080) wide field of view, complete $uv$ coverage in 12 hour tracks, and well-characterized antennas with prime focus receivers are uniquely well-suited to studies of the cold neutral, warm neutral, and warm magnetoionized phases of the ISM. The renewed Synthesis Telescope, with simultaneous $400-1800 \MHz$ continuum Stokes $I$, $Q$, and $U$ observations as well as simultaneous coverage of \Hi, RRLs, and OH over a velocity range sufficiently wide to simultaneously observe disk and high velocity gas, will transform our understanding of magnetized gas in the Galactic ISM. The renewed ST provides an opportunity for Canadian scientific leadership in Galactic ISM science at an excellent radio quiet site at a relatively small capital cost of $\sim\$3$M. The ST will also provide a platform for development of radio astronomy technologies and techniques; phased array feeds on the ST would provide opportunities for the next step change in wide-field spectral line science, enabling sensitive, arcminute coverage of the northern high velocity and high latitude sky.

ISM science requires {\bf large single dish radio telescopes} both for sensitivity and for zero spacing information for interferometers; a large, steerable single dish capability should be maintained. The {\bf Green Bank Telescope} is the only telescope which can provide sensitivity to the molecular lines in the crucial $70-116$~GHz window, with no other options on the horizon. For continuum emission, the {\bf Effelsberg 100~m Telescope} combined with the {\bf Galt Telescope} are the best available instruments to provide zero-spacing information for the ST, while the GBT and Galt Telescope provide zero spacing capability for RRLs.

{\bf Balloon-borne telescopes} operate above 99.5$\%$ of the Earth's atmosphere and therefore have a huge sensitivity advantage over comparable ground-based submm telescopes, providing an exciting platform for studies of the Milky Way in the coming decade.  Building on the success of BLASTPol, the next generation BLAST polarimeter {\bf BLAST-TNG}, will study magnetic fields and dust properties in dozens of regions including GMCs, \Hii\ regions, local regions of the CNM, and nearby galaxies.
BLAST-TNG is scheduled for a first launch from Antarctica in December 2019.  Future balloon-borne polarimeters would benefit from access to superpressure balloons being developed by NASA, which can stay aloft for up to 100 days.  Superpressure balloons can also launch from mid-latitude locations giving access to much larger fraction of the sky.

The DRAO 26 m Telescope was renamed the {\bf John A.\ Galt Telescope} in 2014 in honour of the DRAO's first employee and long-time director. The telescope is nearing completion of a major upgrade including the installation of: modern spectral line and pulsar/FRB backends employing GPUs and a high-bandwidth FPGA signal-processing engine designed by the McGill Cosmology Instrumentation Lab; a customized MeerKAT $L$-band ($900-1800 \MHz$) receiver fitted with NRC-developed low-noise amplifiers; optical fibers and helium lines; and the cryogenic infrastructure required to support helium-cooled receivers. The upgraded Galt Telescope will be optimized for Zeeman splitting measurements of \Hi, OH, and $L$-band RRLs, with a 20-fold increase in bandwidth, 18,000-fold in spectral channels, and 125-fold in number of observable spectral lines across the band.

{\bf VLBI} capability is crucial for further progress in a number of areas of ISM science. VLBI parallax measurements of pulsars provide the best constraint on pulsar distances which in turn provide the best constraint on the distribution of free electrons in the ISM. VLBI parallax measurements of masers provide the best tracer of distances to spiral arms and therefore the distribution of all gas in the Galaxy. VLBI is also the next frontier for pulsar scintillometry measurments, probing the smallest scales of turbulence in the ISM. VLBI between the Galt Telescope (using the CHIME backend) and the Algonquin Radio Observatory is now routine, and investment in VLBI outriggers for CHIME, CHORD, and ultimately the ngVLA will ensure this capability in the long term.

\begin{lrptextbox}[How does the proposed initiative result in fundamental or transformational advances in our understanding of the Universe?]

The ISM mediates star formation and the evolution of galaxies. Therefore, understanding the physical processes which control the flow of energy and matter in the ISM is an essential piece to understanding how galaxies turn gas into stars (see \S~\ref{sec:science}). Moreover, the ISM is a foreground for most other astronomical and cosmological observations through effects including dust extinction and diffuse radio continuum, microwave, and infrared emission; in the era of precision cosmology, accurate understanding of these foregrounds is essential.

\end{lrptextbox}

\begin{lrptextbox}[What are the main scientific risks and how will they be mitigated?]

A significant challenge studying the ISM is piecing together a complete understanding over the wide range of scales and physical processes involved. Every major new observational facility has produced a new vision of the ISM that is more dynamic with more multi-phase interactions than the view before. Although we cannot understand any one phase in isolation, the observational and theoretical programs described here make progress with observations that utilize the wide bandwidths that are now possible in the radio to build data sets which incorporate the diverse constituents of the ISM.

\end{lrptextbox}

\begin{lrptextbox}[Is there the expectation of and capacity for Canadian scientific, technical or strategic leadership?] 

Canada has played a leadership role in ISM science and in designing, building, and operating the necessary facilities for decades. The Canadian Galactic Plane Survey revolutionized our understanding of the ISM, particularly in measuring the interplay between cold and warm atomic gas, star formation, and magnetic fields. Canadians led the ISM science work with {\em Planck}. Canadians are leaders of the POSSUM and GMIMS investigations of magnetoionized gas. The WIDAR correlator on the VLA is Canadian. Looking forward, all of the current and proposed major facilities at DRAO, including CHIME, CHORD, the ST, and the Galt Telescope, will have globally unique capabilities which are directly relevant to ISM science with Canadian leadership. Canadians are leaders in our understanding of dust properties in dense and diffuse clouds using {\em Herschel}, BLASTPol, and BLAST-TNG.

\end{lrptextbox}

\begin{lrptextbox}[Is there support from, involvement from, and coordination within the relevant Canadian community and more broadly?] 

ISM science in Canada is and has been generally organized in collaborations focused around telescopes, including the CGPS, POSSUM, GASKAP, GMIMS, BLASTPol, and {\em Planck}. These collaborations have effectively coordinated their effort. Canadian ISM astronomers are embedded within cosmological experiment teams including {\em Planck} and CHIME and are involved in the development of CHORD.

\end{lrptextbox}

\begin{lrptextbox}[Will this program position Canadian astronomy for future opportunities and returns in 2020-2030 or beyond 2030?] 

The proposed work here includes a number of relatively short-term
and low-cost efforts like the upgrades to DRAO telescopes and continued balloon-borne submm astronomy. The testbed nature of the DRAO makes it able to respond new opportunities as they arise and facilitates the development of new ideas from universities.

\end{lrptextbox}

\begin{lrptextbox}[In what ways is the cost-benefit ratio, including existing investments and future operating costs, favourable?] 

The major new instruments whose primary science mission is related to the ISM -- the Synthesis Telescope, the Galt Telescope, and balloon-borne experiments -- are all relatively inexpensive (each $\lesssim \$3$M in capital cost) upgrades of existing facilities. ISM science also benefits enormously from higher-cost observatories with broad applications (the Green Bank Telescope, ALMA, SKA, and ngVLA) and from experiments with other primary science applications ({\em Planck} and CHIME in the past decade, and CHORD in the 2020s).

\end{lrptextbox}

\begin{lrptextbox}[What are the main programmatic risks
and how will they be mitigated?] 

The main programmatic risks are losing access to the necessary facilities and capabilities, especially computational facilities (\S~\ref{sec:resources}). The facilities which are largely for ISM science (the ST, the Galt Telescope, BLASTPol) are all relatively inexpensive both in capital and operating costs. The science goals also rely on partnerships with major facilities (ASKAP, SKA, ngVLA, ALMA) and experiments primarily built for other purposes but which must understand the ISM as a foreground (CHIME, CHORD). The major frontier in ISM radio astronomy is broadband observations, both for continuum science and for observing many spectra lines -- such as RRLs -- simultaneously. As these frequencies are mostly outside of protected bands, protecting the RFI environment is crucial. Canada's national radio observatory, DRAO, is among the most radio quiet observatories which host working astronomers and engineers (including students) in the world. The Canadian astronomy community should work with governments and other stakeholders to ensure that this continues.

\end{lrptextbox}

\begin{lrptextbox}[Does the proposed initiative offer specific tangible benefits to Canadians, including but not limited to interdisciplinary research, industry opportunities, HQP training,
EDI,
outreach or education?] 

Many of the important ISM projects in Canada use relatively small telescopes that provide hands-on training opportunities for HQP both in science and instrumentation (e.g. the ST, BLASTPol, CHIME). Within Physics and Astronomy, ISM science is inherently multidisciplinary because the ISM is a foreground for cosmological, extragalactic and stellar observations, and because the ISM provides a unique laboratory for high-energy physics. Furthermore, the spectacular wide-field images make ISM science well-suited to public outreach.

\end{lrptextbox}

\setlength{\bibsep}{0.0pt}

\bibliography{bibliography,bib_planck} 

\end{document}